\documentclass{ws-ijmpd}
\usepackage[super,compress]{cite}
\usepackage{graphicx}
\usepackage{epstopdf}
\usepackage{tensor}
\usepackage{undertilde}

\begin{document}

\markboth{J. P. Morais Gra\c ca, A. de P\'adua Santos, Eug\^enio R. Bezerra de Mello and V. B. Bezerra}
{Non-Abelian cosmic string in the Starobinsky model of gravity}

%
\catchline{}{}{}{}{}
%

\title{Non-Abelian cosmic string in the Starobinsky model of gravity}

\author{J. P. Morais Gra\c ca$^{1}$, A. de P\'adua Santos$^{2}$, Eug\^enio R. Bezerra de Mello$^{1}$ and V. B. Bezerra$^{1}$}

\address{$^{1}$Universidade Federal da Para\'iba, Caixa Postal 5008\\
 Jo\~ao Pessoa/PB, CEP 58051-970, Brasil}

\address{$^{2}$Universidade Federal Rural de Pernambuco\\
Rua Dom Manoel de Medeiros, s/n, Dois Irmãos \\
Recife/PE, CEP 52171-900, Brasil}

\address{jpmorais@gmail.com \\ padua.santos@gmail.com \\ emello@fisica.ufpb.br \\ valdir@fisica.ufpb.br}

\maketitle

\begin{history}
\received{Day Month Year}
\revised{Day Month Year}
\end{history}

\begin{abstract}
In this paper, we analyze numerically the behaviour of the solutions corresponding to a non-Abelian cosmic string in the framework of the Starobinsky model, i.e. where $f(R)=R + \zeta R^2$. We perform the calculations for both an asymptotically flat and asymptotically (anti)de Sitter spacetimes. We found that the angular deficit generated by the string decreases as the parameter $\zeta$ increases, in the case of a null cosmological constant. For a positive cosmological constant, we found that the cosmic horizon is affected in a non-trivial way by the parameter $\zeta$.
\end{abstract}




\section{Introduction}

Currently, the most accepted model for the origin and evolution of the observed universe is the so-called standard model of cosmology. Its naive formulation, however, presents conceptual and observational problems, such as the flatness and horizon problems. One way to overcome these issues is to allow for the early universe to go through an accelerated expansion phase, known as inflation. Its causes can be associated to a phase transition of a scalar field, and one consequence of such transition is to provide a mechanism for the formation of topological defects, like domain walls, monopoles, cosmic strings, among others \cite{Hindmarsh:1994re,Vilenkin:2000jqa,Nielsen:1973cs,Sokolov,Vilenkin:1981zs,Garfinkle:1985hr,Christensen:1999wb,Brihaye:2000qr,Sakellariadou:2009ev,BezerradeMello:2003ei,Dzhunushaliev:2006kd}.

In the last decade of the past century, a new major observational issue has challenged our understanding about the cosmos. It has been discovered that our universe is expanding in an accelerated rate, in contrast to the believed fact that matter should always be attractive. This unexpected phenomena can be explained, mainly, in two possible ways, namely, introducing a new kind of matter that simulates a kind of cosmological constant, or replacing general relativity for a more general theory of gravity. 

One simple way to modify general relativity is to add new terms to the Einstein-Hilbert action. If we believe that Einstein's gravity is just the leading order of an effective theory, the first natural correction are the quadratic terms, such as $R^2, R_{\mu\nu} R^{\mu\nu}, R_{\mu\nu\alpha\beta}  R^{\mu\nu\alpha\beta}$ and combinations of them. In this paper we will follow this approach and work with the Starobinsky model of gravity, a model that can be recast as an $f(R)$ theory \cite{Sotiriou:2008rp}, when we assume that $f(R) = R + \zeta R^2$ \cite{Starobinsky:1980te}. Our main goal is to analyze how the introduction of this first order correction affects the results already obtained in general relativity. Specifically, we are interested to investigate the influence of the Starobinsky term on the formation of the linear topological defect named non-Abelian vortex.

A similar study has been performed recently in \cite{MoraisGraca:2016ohe,Graca:2016xcu}, where an Abelian model for a cosmic string has been analyzed in both, an extended Starobinsky model and general relativity, and its properties compared. Despite
the fact that the gravitational field far away from the core of a cosmic string vanishes, it generates
an angular deficit that can, in principle, be observed by their astrophysical effects, and the failure to observe such effects can introduce an upper bound on the energy scale associated with the formation of this defect. In \cite{Graca:2016xcu} it was shown, for example, that the introduction of an $R^n$ term in the action will alter such bounds, allowing the phase transition to occur for higher values of the vaccum expectation value (VEV) of the bosonic field, without affecting the regularity of the space-time.  

In this paper, we will perform a similar analysis, but now considering a non-Abelian model for the cosmic string. This step is necessary to verify the behaviour of such modification also on the non-Abelian case. To do so, we will need to solve the full coupled Einstein-scalar-Yang-Mills field differential equations, a task that can be done only by numerical approach. This will allow us examine how the angular deficit vary along all the parameters of the theory. It will also allow us to study the linear energy density of the cosmic string and, also, in the presence of a cosmological constant, how the cosmological horizon vary as we modify the parameter $\zeta$, related with the Starobinsky correction.

This paper in organized in the following manner. In section 2 we will present our model and the field equations we must solve, along with the most import properties of the spacetime generated by the cosmic string. In section 3 we will solve these equations numerically and present our results comparing it with the previous ones about the same model in Einstein's gravity \cite{Santo:2015xma}. Finally, in section 4 we will present our conclusions.

\section{The model}

The action for a gravitating cosmic string in the Starobinsky model, and in the presence of a cosmological constant, $\Lambda$, is given by

\begin{equation}
S = \int d^4x \sqrt{-g} \left( \frac{1}{2\kappa^2}(R + \zeta R^2 - 2\Lambda) + \mathcal{L}_m \right),
\label{action}
\end{equation}
where $\mathcal{L}_m$ stands for the Lagrangian associated with a cosmic string system, $R$ is the Ricci scalar and $\kappa^2 = 8 \pi G$. The case where $\mathcal{L}_m$ stands for the Abelian-Higgs model is usually called the Abelian cosmic string and was already studied in \cite{MoraisGraca:2016ohe,Graca:2016xcu}. In this paper, we are interested in the study of a particular non-Abelian model for a cosmic string, such as \cite{Nielsen:1973cs}

\begin{equation}
\mathcal{L}_m = - \frac{1}{4} F^a_{\mu\nu} F^{\mu\nu a} + \frac{1}{2}(D_\mu \psi^a)^2 + \frac{1}{2}(D_\mu \chi^a)^2 - V(\psi^a, \chi^a), \hspace{10pt}a=1,2,3.
\end{equation}

In this case, the field strength tensor is given by

\begin{equation}
F^a_{\mu\nu} = \partial_\mu A_\nu^a - \partial_\nu A_\mu^a + e \epsilon A_\mu^b A_\nu^c,
\end{equation}
where the Yang-Mills fields $A_\mu^a$ is in the fundamental representation, and both the scalar fields $\psi^a$ and $\chi^a$ are in the adjoint representation of the $SU(2)$ symmetry group. The covariant derivatives of the scalar fields are given by

\begin{equation}
D_\mu \psi^a = \partial_\mu \psi^a + e \epsilon^{abc} A_\mu^b \psi^c,
\end{equation}
and

\begin{equation}
D_\mu \chi^a = \partial_\mu \chi^a + e \epsilon^{abc} A_\mu^b \chi^c,
\end{equation}
where the latin indices $(a, b, c...)$ run from $1$ to $3$.

We will consider the same potential as considered in \cite{Santo:2015xma},

\begin{eqnarray}
V(\psi^a, \chi^a) &=& \frac{\lambda_1}{4}[(\psi^a)^2 - \eta_1^2]^2 + \frac{\lambda_2}{4}[(\chi^a)^2 -\eta_2^2]^2 
\nonumber
\\
&+& \frac{\lambda_3}{2}[(\psi^a)^2 - \eta_1^2][(\chi^a)^2-\eta_2^2],
\end{eqnarray}
where the parameters $\lambda_1$ and $\lambda_2$ are the self-coupling constants for the scalar fields, and $\lambda_3$ is the coupling constant between the two scalar sectors. The parameters $\eta_1$ and $\eta_2$ correspond to the scale of energy where the symmetry is spontaneously broken. To assure that both scalar fields assume a non-trivial vacuum expectation value at infinity, the parameters $\lambda_1, \lambda_2$ and $\lambda_3$ must assume values such that $\lambda_1 \lambda_2 - \lambda_3^2 > 0$ \cite{Santo:2015xma}.

\subsection{The Ansatz}

Due to the cylindrical symmetry of the model, we can choose the line element for the metric as

\begin{equation}
ds^2 = N^2(\rho) dt^2 - d\rho^2 - L^2(\rho) d\phi^2 - K^2(\rho) dz^2,
\label{metric}
\end{equation}
where, for this particular model, we can set $K(\rho) = N(\rho)$, since there is a symmetry between the $(tt)$ and $(zz)$ components of the energy-momentum tensor, as we will see. 

For the scalar fields, we will use the \textit{ansatz} \cite{deVega:1976rt}

\begin{equation}
\psi^a(\rho) = f(\rho) \left( \begin{array}{c} \text{cos}\hspace{4pt} \phi \\ \text{sin}\hspace{4pt} \phi \\ 0 \end{array} \right)
\end{equation}
and
\begin{equation}
\chi^a(\rho) = g(\rho) \left( \begin{array}{c} -\text{sin} \hspace{4pt} \phi \\ \text{cos} \hspace{4pt} \phi \\ 0 \end{array} \right).
\end{equation}

From which we verify that $\psi^a \psi^a = f^2$, $\chi^a \chi^a = g^2$ and $\psi^a \chi^a = 0$. The ansatz for the gauge fields are

\begin{equation}
\vec{A}^a (\rho) = \hat{\phi} \left( \frac{1 - H(\rho)}{e\rho} \right) \delta_{a,3} = - \hat{\phi} \frac{A(\rho)}{e\rho} \delta_{a,3},
\end{equation}
and

\begin{equation}
A_t^a(\rho) = 0, \hspace{10pt}a=1,2,3.
\end{equation}

Indeed, our ansatz for the gauge field in this non-Abelian model is similar to the usual ansatz for the Abelian model, but now localized in the third component of the triplet. This is the same ansatz used by Nielsen and Olesen in their original paper on cosmic strings \cite{Nielsen:1973cs}. Despite the fact that the gauge fields are not so different from the Abelian case, this model is not trivial, since the gauge fields are affected by its interaction with the scalar fields. 

We also used the first winding number is all our considerations. The main reason is that the winding number is more important for the structure of the cosmic string than for the spacetime generated by the string. For this reason, and also due to the large number of parameters already available for this model, we decided to not consider a winding number other than unity.
\subsection{Equations of motion}

The equations of motion are obtained varying the action (\ref{action}) with respect to the matter fields and the metric. Before we present our results, we will do some function and parameter redefinitions with the purpose to deal only with dimensionless fields and parameters. 

Let us define a characteristic length scale given by $x = \sqrt{\lambda_1} \eta_1 \rho$. Based on such definition, we can now work with dimensionless functions 

\begin{equation}
f(\rho) = \eta_1 X(x), \hspace{10pt} g(\rho) = \eta_1 Y(x) \hspace{10pt} 
\end{equation}
for the scalar fields, and

\begin{equation}
L(x) = \sqrt{\lambda_1} \eta_1 L(\rho), \hspace{10pt}R(x)=\sqrt{\lambda_1} \eta_1 R(r)
\end{equation}
for the metric $L(x)$ function and the Ricci scalar curvature. The (almost) free parameters of the theory can also be redefined as

\begin{equation}
\hspace{-50pt} \alpha = \frac{e^2}{\lambda_1}, \hspace{5pt} q = \frac{\eta_2}{\eta_1}, \hspace{5pt} \beta_i^2 = \frac{\lambda_i}{\lambda_1} \hspace{5pt}i=1,2,3, \hspace{5pt}\gamma = \kappa^2\eta_1^2\hspace{5pt}\text{and}\hspace{5pt}\xi = \zeta \lambda_1 \eta_1^2, \hspace{5pt} \bar{\Lambda} = \frac{\Lambda}{\eta_1^2 \lambda_1}.
\end{equation}

The field equations for the scalar and gauge fields are given by \cite{Santo:2015xma},

\begin{eqnarray}
\label{eqnX}
\frac{(N^2LX')'}{N^2L} = X\left[X^2 - 1 + \beta_3^2(Y^2-q^2)+\frac{H^2}{L^2}\right]
\\
\label{eqnY}
\frac{(N^2LY')'}{N^2L} = Y\left[\beta_3^2(X^2-1)+\beta_2^2(Y^2-q^2)+\frac{H^2}{L^2}\right]
\\
\label{eqnH}
\frac{L}{N^2}\left(\frac{N^2H'}{L}\right)' = \alpha (X^2 + Y^2)H,
\end{eqnarray}
where the prime $'$ means derivative with respect to the redefined variable $x$.

The energy-momentum tensor for the matter fields is calculated as usual, $T^{\mu\nu} = \frac{2}{\sqrt{g}} \frac{\delta S_{matter}}{\delta g_{\mu\nu}}$, and is given by

\begin{eqnarray}
T_t^t = T_z^z = \eta_1^4 \lambda_1 [\epsilon_1 + \epsilon_2 + \epsilon_3 + \epsilon_4 + \epsilon_5 + \epsilon_6 + \epsilon_7 + \epsilon_8]
\\
T_\rho^\rho = \eta_1^4 \lambda_1 [- \epsilon_1 - \epsilon_2 + \epsilon_3 - \epsilon_4 + \epsilon_5 + \epsilon_6 + \epsilon_7 + \epsilon_8]
\\
T_\phi^\phi = \eta_1^4 \lambda_1 [- \epsilon_1 + \epsilon_2 - \epsilon_3 + \epsilon_4 - \epsilon_5 + \epsilon_6 + \epsilon_7 + \epsilon_8]
\end{eqnarray}
where

\begin{eqnarray}
\epsilon_1 = \frac{1}{2} \frac{H'^2}{\alpha L^2}, \hspace{8pt} \epsilon_2 = \frac{1}{2} {X'^2}, \hspace{8pt} \epsilon_3 = \frac{1}{2} \frac{X^2 H^2}{L^2}, \hspace{8pt} \epsilon_4 = \frac{1}{2} Y'^2,\hspace{8pt} \epsilon_5 = \frac{1}{2} \frac{Y^2 H^2}{L^2}
\\
\epsilon_6 = \frac{1}{4}(X^2-1)^2, \hspace{8pt} \epsilon_7 = \frac{1}{4} \beta_2^2 (Y^2 - q^2)^2, \hspace{8pt} \epsilon_8 = \frac{1}{2} \beta_3^2 (X^2-1)(Y^2-q^2). 
\end{eqnarray}

In the original parameters, the gravitational field equations are given by

\begin{equation}
G_{\mu\nu}(1 + 2 \zeta R) + \frac{1}{2} g_{\mu\nu}( \zeta  R^2 + 2 \Lambda) - 2\zeta (\nabla_\mu \nabla_\nu - g_{\mu\nu} \Box) R = - \kappa^2 T_\mu\nu,
\label{fRequations}
\end{equation}
where $G_{\mu\nu}$ is the Einstein tensor and $\Lambda$ is the cosmological constant. This is a set of fourth-order differential equations in the metric, but we can always decrease the order of the differential equations enlarging the number of independent components. This is the reason why we treat the Ricci scalar as an independent variable. With this trick, we have a set of second-order coupled differential equations. The trace of (\ref{fRequations}) is given by

\begin{equation}
-R + 6 \zeta \Box R + 4 \Lambda = - \kappa^2 T,
\label{eqTrace} 
\end{equation} 
and, far away from the source, we have that the Ricci scalar is given by $R = 4\Lambda$. As we can see, in the absence of a cosmological constant, the Starobinsky model imposes an asymptotically flat solution. This is not the case for an arbitrary $f(R)$ polynomial such as $f(R) = R + \zeta_2 R^2 + \zeta_m R^m$.

Substituting the redefined parameters and functions at (\ref{fRequations}), and after some algebraic manipulations, the second-order field equations that the metric functions and the Ricci scalar should obey are given by

\begin{eqnarray}
\label{eqqR}
R'' = \frac{1}{24} \frac{1}{N^2 L^2 \alpha \xi}[6 \xi \alpha L R(8 N' N L' + 4 N'^2 L + R L N^2) \\
\nonumber
- \gamma \alpha N^2 [L^2(X^2-1)^2 + L^2(Y^2 - q^2)^2] + 10L^2(X'^2 + Y'^2) \\
\nonumber
- 2H^2(X^2 + Y^2) + 2 \beta_3^2 L^2(X^2-1)(Y^2-q^2)] - 6 \gamma N^2 H'^2 \\
\nonumber
+ 4 \alpha(6 N' L N L' + 3 N'^2 L^2 + R N^2 L^2) - 4 \bar{\Lambda} \alpha L^2 N^2],
\end{eqnarray}
\begin{eqnarray}
\label{eqqN}
N'' =  -\frac{1}{24} \frac{1}{\alpha N L^2 (1 + 2 \xi R)}[6 \xi \alpha L (4 R L N'^2+ R^2 L N^2- 4 R' L' N^2) 
\\
\nonumber
- \gamma \alpha N^2 [L^2(X^2-1)^2 + L^2(Y^2 - q^2)^2] - 2L^2(X'^2 + Y'^2) \\
\nonumber
+ 10H^2(X^2 + Y^2) + 2 \beta_3^2 L^2(X^2-1)(Y^2-q^2)] - 6 \gamma N^2 H'^2 \\
\nonumber
+ 4 \alpha L^2 (3 N'^2 + R N^2) - 4 \bar{\Lambda} \alpha L^2 N^2],
\end{eqnarray}
and 
\begin{eqnarray}
\label{eqqL}
L'' = \frac{1}{24} \frac{1}{L \alpha N^2(1 + 2 \xi R)}[6 \xi \alpha L(4 N'^2 L R - R^2 N^2 L - 8 N' N L' R \\
\nonumber
+ \gamma \alpha N^2 [L^2(X^2-1)^2 + L^2(Y^2 - q^2)^2] - 2L^2(X'^2 + Y'^2) \\
\nonumber
- 14H^2(X^2 + Y^2) + 2 \beta_3^2 L^2(X^2-1)(Y^2-q^2)] - 18 \gamma N^2 H'^2 \\
\nonumber
+ 4 \alpha L(3 N'^2 L - R N^2 L - 6 N' N L') + 4\bar{\Lambda} \alpha N^2 L^2],
\end{eqnarray}
where one more time the prime $'$ means derivative with respect to the redefined coordinate $x$. We must now solve the coupled equations (\ref{eqnX}-\ref{eqnH}) and (\ref{eqqR}-\ref{eqqL}), with the appropriate boundary conditions, that we will discuss in what follows.

\subsection{Boundary conditions}

Here we will provide the boundary conditions that the metric functions and the matter fields must obey. The latter are imposed to achieve both the regularity of the functions at the origin, and also the requirement that the fields approaches its vacuum expected values at infinity. Therefore, we must have

\begin{equation}
H(0) = 1, \hspace{10pt} H(x_{max}) = 0,
\label{bc1}
\end{equation}
where $x_{max}$ is infinity for vanishing or negative cosmological constant, and corresponds to the first zero of the function $N(x)$, for a positive cosmological constant. We must also impose

\begin{equation}
X(0) = 0, \hspace{10pt} X(\infty) = 1, \hspace{10pt}Y(0) = 0, \hspace{10pt} Y(\infty) = \frac{\eta_2}{\eta_1} = q.
\label{bc2}
\end{equation}

For the metric functions, the regularity at the origin is obtained by the following boundary conditions,

\begin{equation}
N(0) = 1, \hspace{10pt} N'(0)=0, \hspace{10pt} L(0) = 0, \hspace{10pt} L'(0) = 1.
\label{bc3}
\end{equation}  

For the Ricci scalar field, the boundary conditions at infinity must satisfy the equation (\ref{eqTrace}), which means that  

\begin{equation}
R(\infty) = 4\bar{\Lambda} \hspace{10pt}\text{and}\hspace{10pt}R'(\infty) = 0.
\label{bc4}
\end{equation}

We will treat separately the cases where $\bar{\Lambda} = 0$ and $\bar{\Lambda} \neq 0$, since the geometry generated by the cosmic string is not the same. In an asymptotically flat spacetime, we can define an angular deficit generated by the string, but the same is not possible in an asymptotically (anti) de Sitter spacetime.

\subsection{Asymptotically flat spacetime and angular deficit}

The curvature generated by the cosmic string decays fast as we move away from its core. However, the same string induces a non-trivial topology that can be probed by an angular deficit around its axis. Asymptotically, the metric functions are given by

\begin{equation}
N(x \rightarrow \infty) = a,
\end{equation}

\begin{equation}
L(x \rightarrow \infty) = bx + c, \hspace{10pt} b \geq 0,
\end{equation}
where $a$, $b$ and $c$ are constants that depends, in principle, of all parameters of the theory. In the absence of sources, our spacetime is a Minkowski one, and we can set $a=1$, $b=1$ and $c=0$.

The angular deficit generated by the cosmic string can be parametrized as

\begin{equation}
\Delta = 2 \pi (1 - L'(\infty)), 
\end{equation}
where, usually, $L'(\infty)$ must be calculated numerically. For the Abelian string in Einstein's gravity, the angular deficit is strongly affected by the value of the vacuum expected valued of the scalar field, as measured by the parameter $\gamma$. We can, in principle, increase the value of the parameter $\gamma$ until $\Delta$ reaches $2 \pi$. The value for the parameter $\gamma$ when $\Delta = 2\pi$ is called critical $\gamma$, and denoted by $\gamma_{cr}$. This value can be used to estimate a maximum upper bound for the VeV of the scalar field. 

The way as the angular deficit changes as we vary the other parameters of the non-Abelian Higgs model was showed in \cite{Santo:2015xma}, where a special attention has been drawn on the parameter $\beta_3$, that represents the coupling between the two scalar fields of the model. In \cite{MoraisGraca:2016ohe}, it has been studied how the angular deficit changes as we vary the parameter $\xi$, related with the squared Ricci term correction, using an Abelian model to represent the cosmic string. It has been shown that as the parameter $\xi$ grows, the angular deficit becomes smaller, allowing a larger VeV for the scalar field before we reach critical value of $\gamma$. This means that, if the Starobinsky model is a good effective field theory for gravity, will be harder to detect gravitational lensing effects due to cosmic strings, and the upper bounds for the VeV due to the missing detection of such effect must be reviewed.

Our first aim in this paper is to verity if the same behaviour happens also in the non-Abelian cosmic string, and compare both results presented above.

\subsection{The linear energy density of the cosmic string}

The energy per unity of length of the cosmic string is defined as

\begin{equation}
\epsilon = \int \sqrt{^{(3)}g}T^t_t d\rho d\phi,
\end{equation} 
with $\sqrt{^{(3)}g}$ being the determinant of the $(2+1)$-dimensional metric (\ref{metric}) with the $z$-axis excluded.  

\subsection{Non-asymptotically flat spacetime and the cosmic horizon}

In an asymptotically (anti-)de Sitter spacetime it is not possible to define an angular deficit as we did for an asymptotically flat spacetime. But in the case of de Sitter spacetime, it it possible to define a cosmological horizon beyond which all events are not causally connected, and we can use the size of such cosmological horizon to parametrize the effect of the non-Abelian string on the geometry.

In vacuum, an axially symmetric spacetime that obeys Einstein's equation with a cosmological constant $\bar{\Lambda}$, can be described by the the line element (\ref{metric}) with \cite{Linet:1986sr}

\begin{equation}
N(x) = \bigg\{ \begin{array}{ll} \text{cos}^{2/3} (\sqrt{3|\bar{\Lambda}|}\frac{x}{2})\\ \text{cosh}^{2/3} (\sqrt{3|\bar{\Lambda}|}\frac{x}{2})\end{array}
\hspace{10pt} \begin{array}{ll} \bar{\Lambda} > 0 \\ \bar{\Lambda} < 0 \end{array}
\end{equation}

and
\begin{equation}
\hspace{-30pt} L(x) = \frac{2^{2/3}}{\sqrt{3|\bar{\Lambda}|}} \left[ 
\bigg\{ \begin{array}{ll} \text{sin}^{1/3} (\sqrt{3|\bar{\Lambda}|}x)\\ \text{sinh}^{1/3} (\sqrt{3|\bar{\Lambda}|}x) \end{array}
\right] \left[ 
\bigg\{ \begin{array}{ll} \text{tan}^{2/3} (\sqrt{3|\bar{\Lambda}|} \frac{x}{2})\\ \text{tanh}^{2/3} (\sqrt{3|\bar{\Lambda}|} \frac{x}{2}) \end{array}\right] \hspace{10pt} \begin{array}{ll} \bar{\Lambda} > 0 \\ \bar{\Lambda} < 0 \end{array},
\end{equation}
where the upper solutions are for de-Sitter spacetime, and the bottom (hyperbolic) solutions are for anti-de Sitter spacetime.

The above solutions can be seen as the spacetime generated by a cosmic string when $\gamma = 0$. In this regime, the matter fields decouple from the metric. The cosmological horizon is identified with the first zero of the function $N(x)$ for de Sitter, given by $x_{ch} = \pi / \sqrt{3 \bar{\Lambda}}$. As we increase the value of the parameter $\gamma$, the value of $x_{ch}$ decreases \cite{Santos:2015mja}. In \cite{Graca:2016xcu} it has been show that in the Starobinsky model of gravity, the value of $x_{ch}$ is not affected by the parameter $\xi$, as long as $\gamma = 0$. In other words, the cosmological horizon is the same in the absence of sources other than a cosmological constant, and depends only on $\bar{\Lambda}$. In the presence of a cosmic string, the major effect of the $\xi R^2$ term is to increase the value of the cosmological horizon. This happens because, in the Starobinsky model of gravity, the gravitational effects due to the cosmic string decreases as the parameter $\xi$ increases. The mentioned study has been performed for the Abelian Higgs model, and one of the main objectives of this paper is to verify if the same effect appears also in the framework of the non-Abelian model for the string. This can help us to understand the effect of a squared curvature term for a more general axially symmetric source. 

In this paper we will also briefly study the effect of the Starobinsky model in the anti-de Sitter spacetime. Such study has not been performed in \cite{Graca:2016xcu}.

\section{Numerical results}

We must numerically solve the equations (\ref{eqnX}-\ref{eqnH}) and (\ref{eqqR}-\ref{eqqL}), together with the proper boundary conditions (\ref{bc1}-\ref{bc4}). To do this we used a well-known ODE solver named COLSYS \cite{Archer1,Archer2}, that implements a Newton-Raphson algorithm to solve the non-linear system of equations obtained after the discretization of the space. Starting with some know profile for the metric and field functions, we slowly increase any of the continuous parameters of the theory to get our desired results. In principle, this method is valid as long as we avoid to reach values in the space of parameters where the functions ceases to be regular. Using this method, the relative errors for the obtained functions are of the order of $10^{-8}$ to $10^{-10}$, and sometimes even better.

The limit $\xi \rightarrow 0$ corresponds to general relativity. However, we cannot use $\xi = 0$, because the equation for the Ricci scalar is not well-defined is this limit. In general relativity, the Ricci scalar obeys an algebraic equations with respect to the energy-momentum tensor, and not a differential equation. But we can choose $\xi$ as small as we want, and in the limit of a small parameter, the results are comparable with the ones obtained in general relativity. We can then slowly increase the parameter $\xi$ to move from general relativity to the Starobinsky model of gravity.  

\subsection{Asymptotically flat spacetime}

\begin{figure}[htb]
\centering
\begin{tabular}{@{}cc@{}}
\includegraphics[scale=0.8]{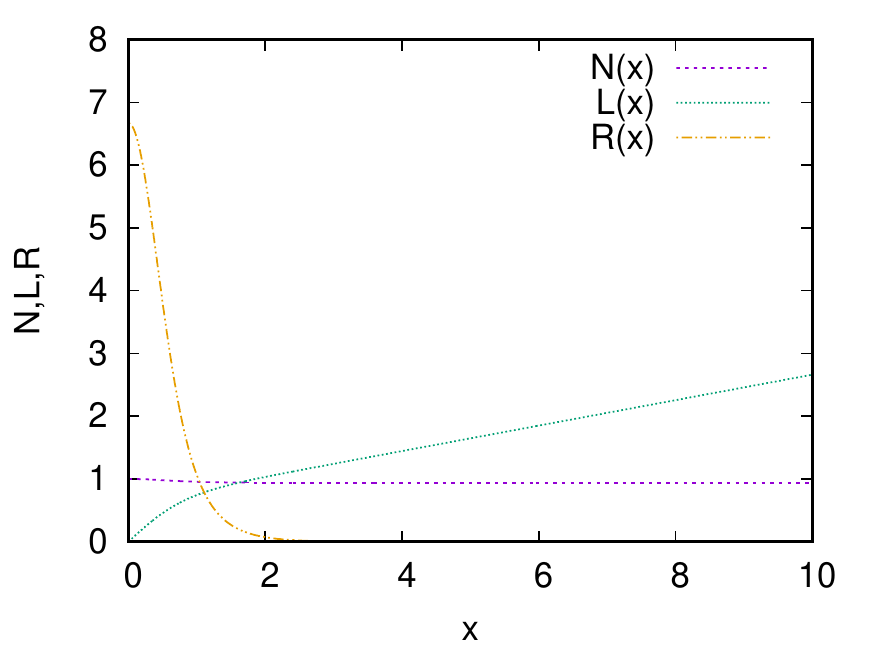} &
\includegraphics[scale=0.8]{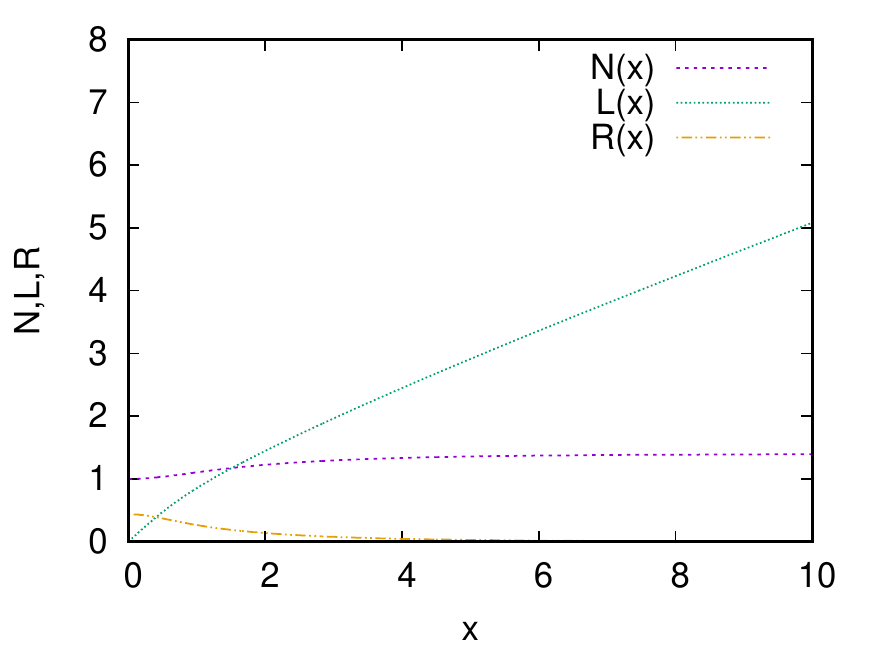}
\end{tabular}
\caption{Profile of the metric functions and the Ricci scalar for the non-Abelian string with parameters $\alpha=1.0$, $\gamma = 0.6$, $\beta_2 = 2.0$, $\beta_3 = 1.0$ and $q=1$. At left, we have $\xi = 0.001$ and at right $\xi = 1.0$.}
\label{fig:nonAbelianprofile}
\end{figure}

In an asymptotically flat spacetime, our main interest will concern with the metric function $L(x)$, whose inclination with respect to the $x$ axis measures the angular deficit. We will be also interested in the curvature measured by the Ricci scalar. In figure (\ref{fig:nonAbelianprofile}) we plot the profile of the metric functions, and the scalar curvature, for a particular set of parameters, with $\xi = 0.001$ and $\xi = 1.0$. As we can see, the functions $L(x)$ and $R(x)$ are strongly affected by increasing the parameter $\xi$, or, in other words, when we move from general relativity to the Starobinsky model of gravity. The inclination of the function $L(x)$ in respect to the $x$ axis increases, which means that the angular deficit generated by the cosmic string decreases. In fact, in general relativity the angular deficit (over $2 \pi$) is $0.797$, and when $\xi = 1.0$ the angular deficit decreases to $0.583$.

This kind of behaviour is expected, since it has already been shown to occur for the Abelian cosmic string in the same gravitational model. Our first goal in this section is to compare the strongness of the effect relative to both types of cosmic string.

We should also note that both the metric function $N(x)$ and the Ricci field are affected as we increase the parameter $\xi$. The former is not so important for our study in the case of an asymptotically flat spacetime, but the effect on the latter deserve some attention, since it measures the curvature near the core of the string. We can note that, in the Starobinsky model of gravity, the curvature generated by the string is much weaker than the same system in general relativity. The reason to be so is related to the fact that the quadratic correction is more important at the strong gravity regime.

In the following subsections we will vary the parameters of the theory. Our goal is to look for patterns on how the non-Abelian cosmic string in the Starobinsky model of gravity differs from the same system in general relativity.

\subsubsection{The parameter $\alpha$:}

The parameter $\alpha$ can also be defined as a ratio between the scalar and gauge masses, and it appears in both the Abelian and non-Abelian string models. To study how the angular deficit and energy vary with both the parameters $\alpha$ and $\xi$, we have fixed all the other parameters but those. We have set $\beta_2 = 1.5$, $\beta_3 = 0.1$ and $\gamma = 0.3$, and calculated the angular deficit and the energy density of the string for the values $\xi = 0.01$, $1.0$ and $100.0$. The obtained results are listed in table (\ref{tableVaryAlpha}). 

The first thing we can notice is that the angular deficit decreases as we increases the parameter $\xi$. As we mentioned before, this was already expected, since this behaviour were already obtained for the Abelian cosmic string \cite{MoraisGraca:2016ohe}, and it is natural (although not obvious) that the non-Abelian string follows the same pattern. We can also note that the effect on decreasing the angular deficit as we increase the parameter $\xi$ is stronger as the parameter $\alpha$ becomes weaker, but this is probably due to the fact that the attenuation on the angular deficit is stronger for large values of the angular deficit. 

\begin{figure}[]
\centering
\begin{tabular}{@{}cc@{}}
\includegraphics[scale=0.80]{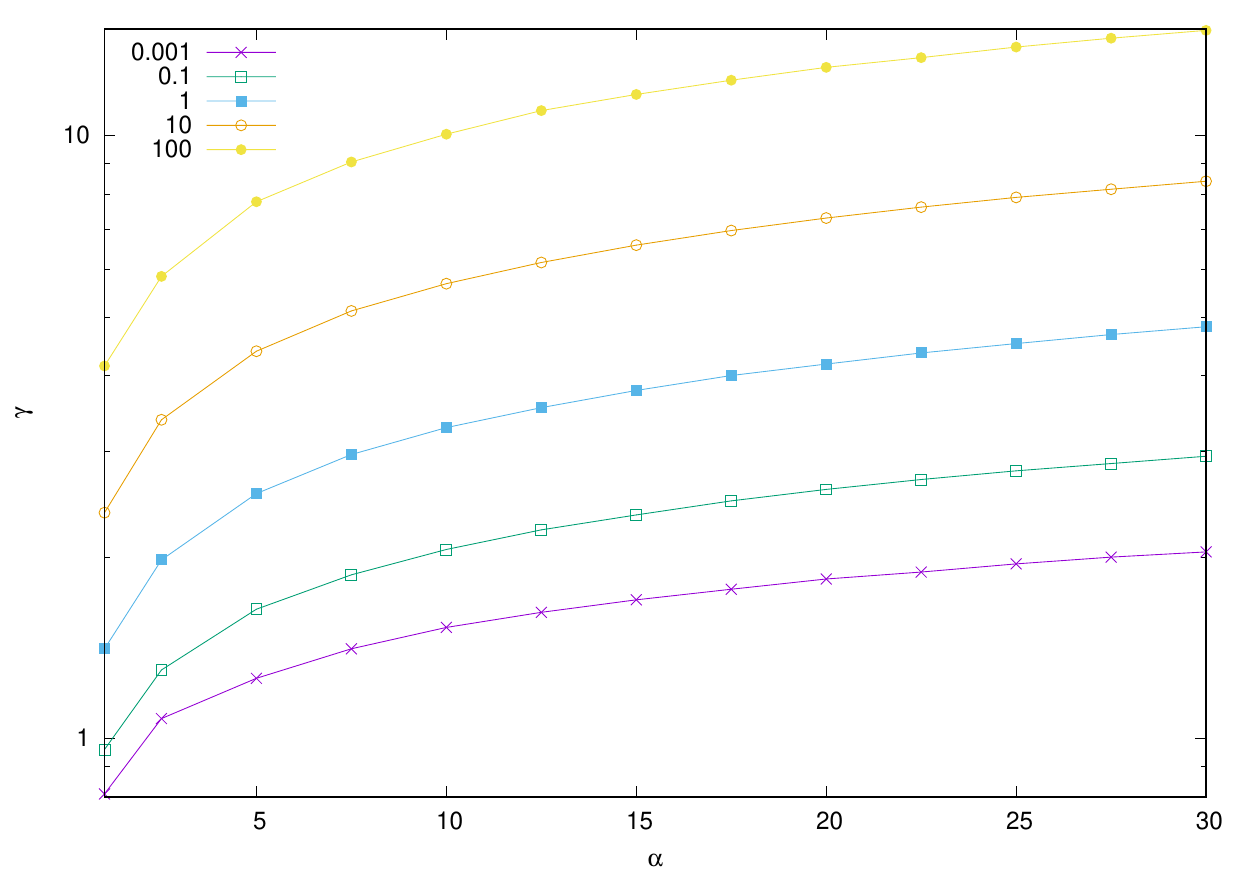}
\end{tabular}
\caption{Values of critical $\gamma$ as a function of the parameter $\alpha$, for several values of the parameter $\xi$. The values used for the string where $\beta_2 = 2.0$ and $\beta_3 = 1.0$.}
\label{fig:RegularProfile}
\end{figure}

\begin{table}[]
\centering
\caption{Values of the linear energy density and angular deficit for $\beta_2 = 1.5$, $\beta_3 = 0.1$ and $\gamma = 0.3$.}
\label{tableVaryAlpha}
\begin{tabular}{p{0.15\textwidth}p{0.15\textwidth}p{0.15\textwidth}p{0.15\textwidth}p{0.15\textwidth}p{0.15\textwidth}}
\hline
 $\alpha, \xi \rightarrow 0$  & Energy  & $\Delta / 2\pi$ \\ \hline
 0.1  & 1.8437 & 0.576 \\
 0.5  & 1.2906 & 0.387 \\
 1.0  & 1.1072 & 0.328 \\
 2.0  & 0.9535 & 0.281 \\
 4.0  & 0.8259 & 0.242 \\
\end{tabular}
\begin{tabular}{p{0.15\textwidth}p{0.15\textwidth}p{0.15\textwidth}p{0.15\textwidth}p{0.15\textwidth}p{0.15\textwidth}}
\hline
 $\alpha, \xi = 1$  & Energy & $\Delta / 2\pi$ \\ \hline
 0.1  & 1.9158 & 0.494 \\
 0.5  & 1.3277 & 0.341 \\
 1.0  & 1.1357 & 0.292 \\
 2.0  & 0.9757 & 0.253 \\
 4.0  & 0.8433 & 0.220 \\
\end{tabular}
\begin{tabular}{p{0.15\textwidth}p{0.15\textwidth}p{0.15\textwidth}p{0.15\textwidth}p{0.15\textwidth}p{0.15\textwidth}}
\hline
 $\alpha, \xi = 100$  & Energy & $\Delta \phi / 2\pi$ \\ \hline
 0.1  & 1.8884 & 0.381 \\
 0.5  & 1.3179 & 0.282 \\
 1.0  & 1.1300 & 0.248 \\
 2.0  & 0.9724 & 0.219 \\
 4.0  & 0.8415 & 0.194 \\
\end{tabular}
\end{table}

In figure (\ref{fig:RegularProfile}) we plot the value for critical $\gamma$ as a function of both $\alpha$ and $\xi$. For all values bellow the curves, the space-time is regular and well-defined. We can note that critical $\gamma$ increases for larger values of the parameter $\alpha$, and also for larger values of the parameter $\xi$. This means that in this polinomial $f(R)$ gravity, the constraints in the $(\gamma, \alpha)$-plane due to the necessity of a regular space-time are relaxed.

\subsubsection{The parameter $\beta_3$}

\begin{figure}[htb]
\centering
\begin{tabular}{@{}cc@{}}
\includegraphics[scale=0.65]{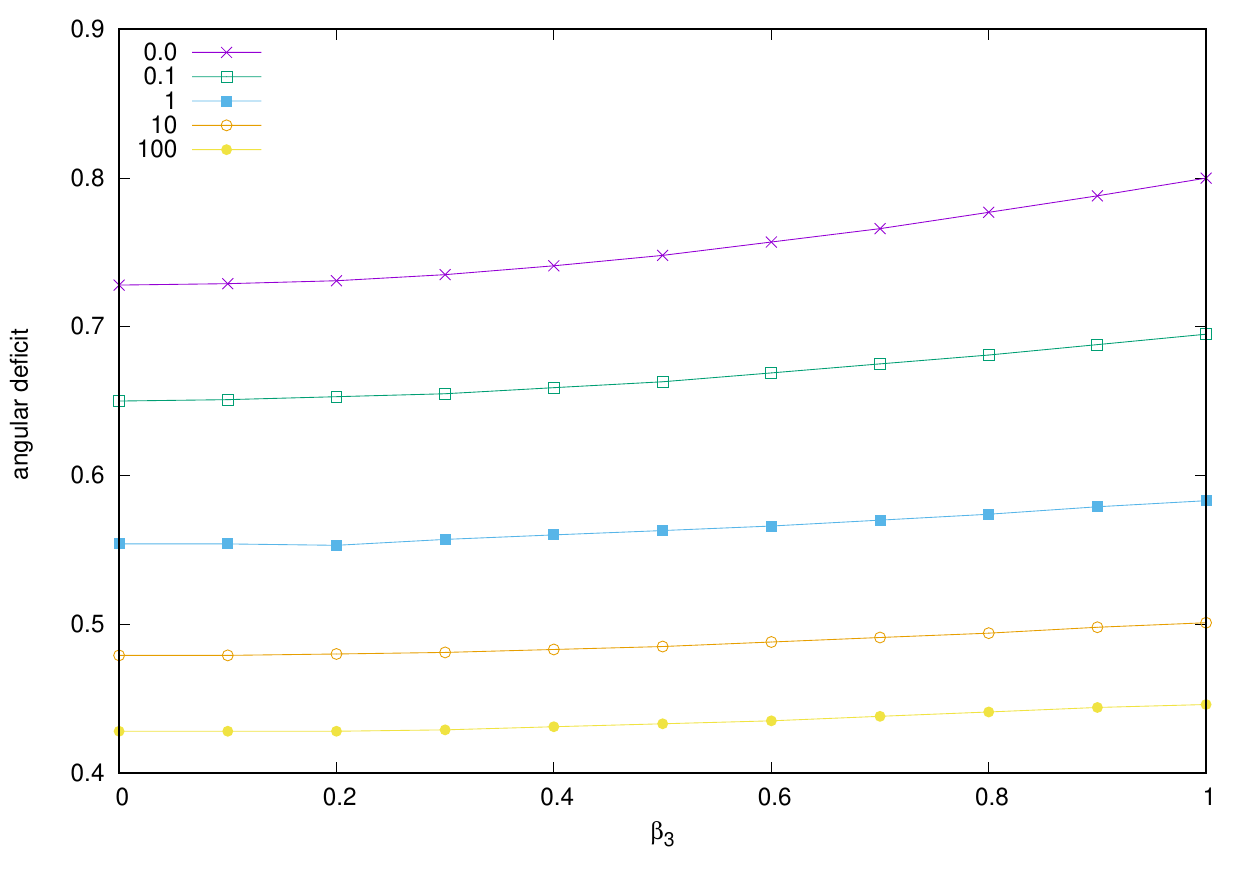} &
\includegraphics[scale=0.65]{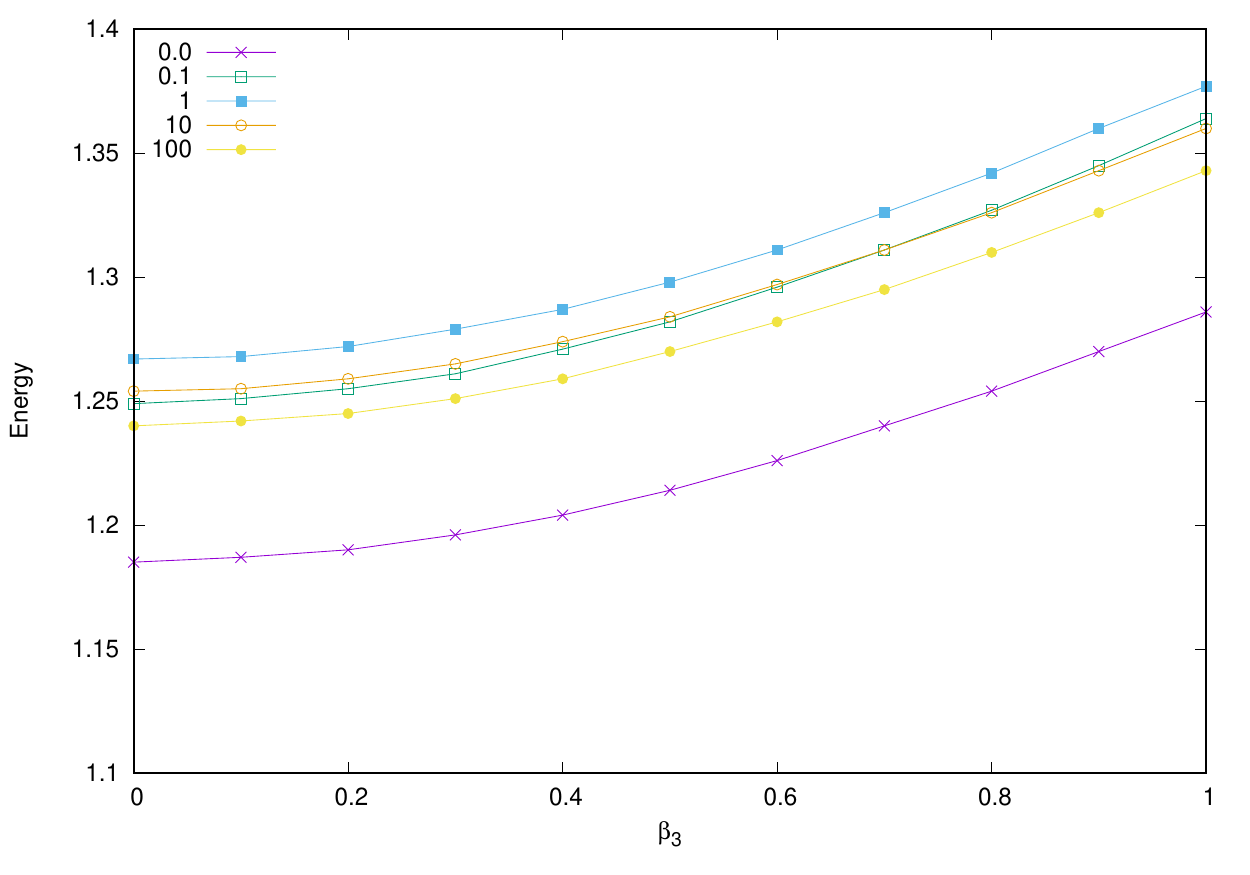}
\end{tabular}
\caption{Angular deficit (over $2 \pi$) generated by the string (left) and energy density of the string (right) as a function of the parameter $\beta_3$, for various values of the parameter $\xi$, for fixed values $\alpha=1.0$, $\beta_2 = 2.0$ and $\gamma = 0.6$.}
\label{fig:beta3Profile}
\end{figure}

The parameter $\beta_3$ stands for the coupling between the two scalar fields in the theory. In\cite{Santo:2015xma} it has been discussed in more details the dependence of the angular deficit and energy density for the non-Abelian cosmic string as functions of the parameter $\beta_3$. Here we are interested to observe how such behaviour changes with the parameter $\xi$, and our obtained results are plotted in figure (\ref{fig:beta3Profile}). 

At left panel, we show the angular deficit over $2 \pi$ ($\Delta / 2 \pi$). As usual, the angular deficit decreases as $\xi$ increases. It is important to note that the curve drawn by the angular deficit as a function of $\beta_3$ is flattened, which means that the parameter $\beta_3$ is more relevant in general relativity than in the Starobinsky model of gravity. This same behaviour could be already notice in Table (\ref{tableVaryAlpha}) for the parameter $\alpha$. 

At right panel, we show how the energy density changes as we increase both the parameters $\beta_3$ and $\xi$. It is interesting to note that the maximum of the energy density is achieved around the value $\xi = 1$. The exact value for the maximum of energy, however, depends on the choice of the parameters of the string, like $\beta_2$ and $\beta_3$.

\subsubsection{The parameter $\gamma$}

\begin{figure}[htb]
\centering
\begin{tabular}{@{}cc@{}}
\includegraphics[scale=0.65]{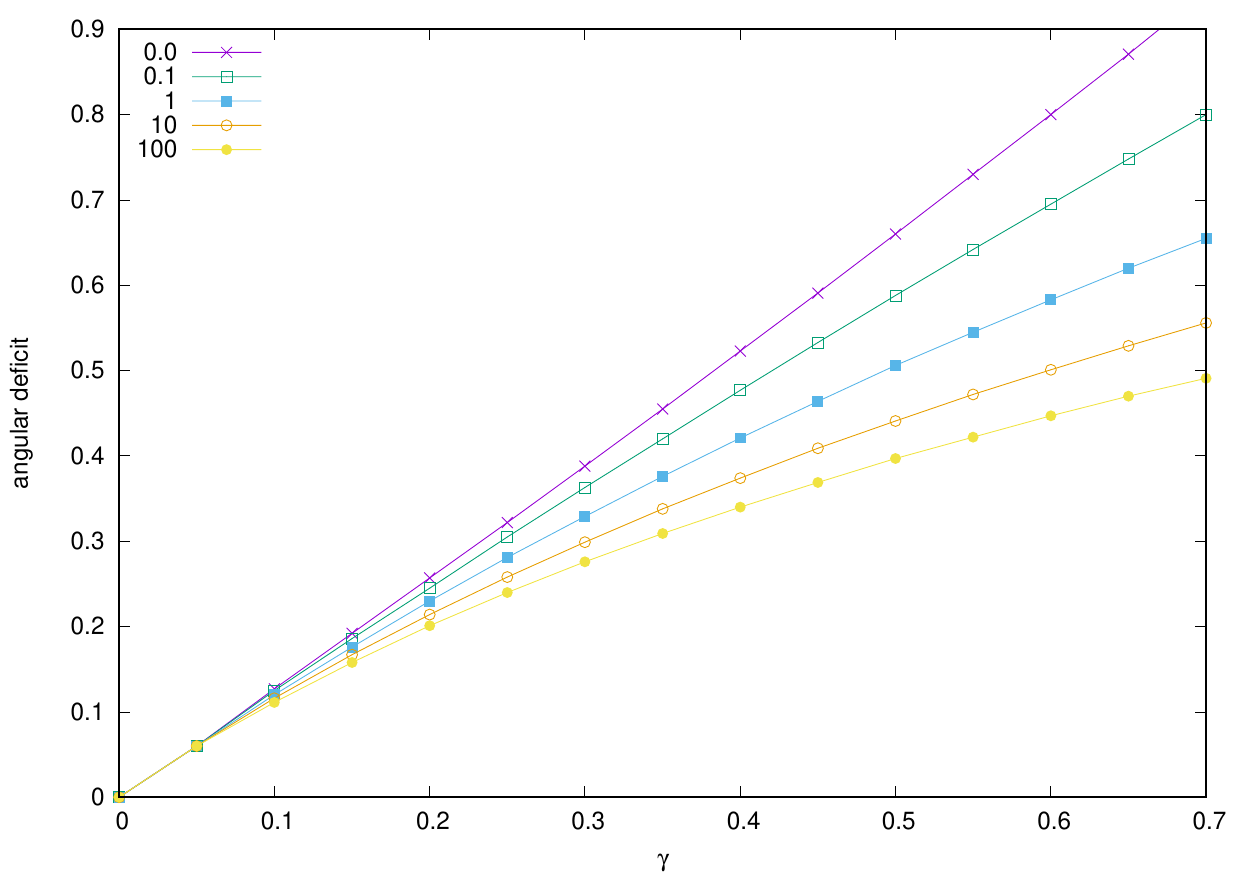} &
\includegraphics[scale=0.65]{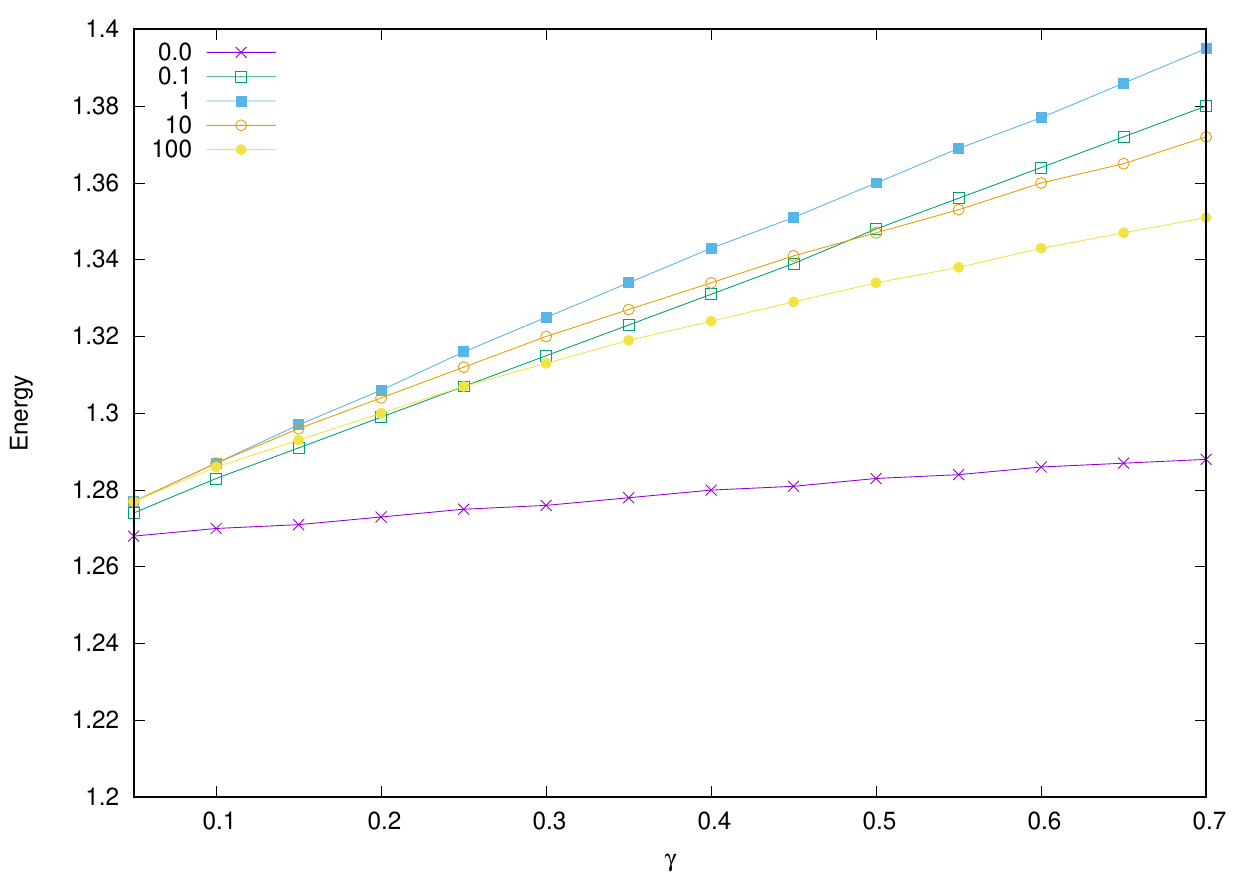}
\end{tabular}
\caption{Angular deficit generated by the string (left), and energy density of the string (right), for several values of the parameter $\xi$. The values used for the cosmic string where $\alpha = 1.0, \beta_2 = 2.0$ and $\beta_3 = 1.0$.}
\label{fig:GammaProfile}
\end{figure}

The parameter $\gamma$ is responsible for the coupling between the string and the gravitational field. If $\gamma$ equals zero, there is no angular deficit.  In figure (\ref{fig:GammaProfile}) we plot the angular deficit and energy as a function of the parameter $\gamma$. For small values of this parameter, the angular deficit curve is the same for any value of $\xi$. However, for large values of $\gamma$, the results obtained by $f(R)$ gravity strongly differs from the one obtained in general relativity. The energy also increases with $\gamma$, however it increases more for a $x_i$ value close to unity. 

\subsection{Cosmic string with a cosmological constant}

As mentioned, in the presence of a positive cosmological constant (de Sitter), the metric has a singularity that can be attributed to a cosmological horizon. Such horizon occurs at the first zero of the function $N(x)$ of the metric, and we will denote it by $x_0$. The value of the horizon changes as we change the parameters of the theory, and it decreases as we increase the parameter $\gamma$. The explanation for such phenomena is straightforward: As we increase the parameter $\gamma$, we are increasing the coupling between the matter and the gravitational field. As the coupling become stronger, gravitational effects become stronger, and it is reasonable to guess that the cosmological horizon will be shorter.

\begin{figure}[htb]
\centering
\begin{tabular}{@{}cc@{}}
\includegraphics[scale=0.65]{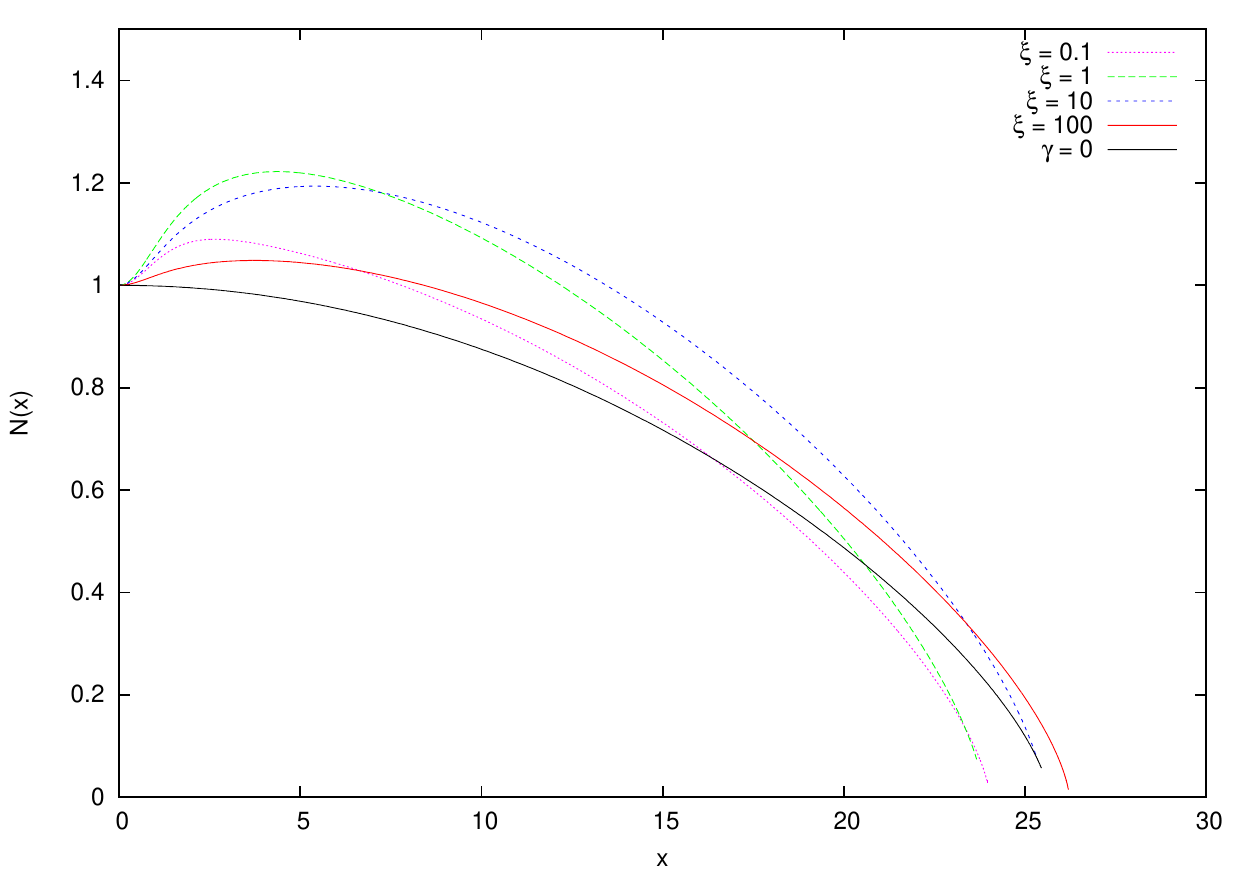} &
\includegraphics[scale=0.65]{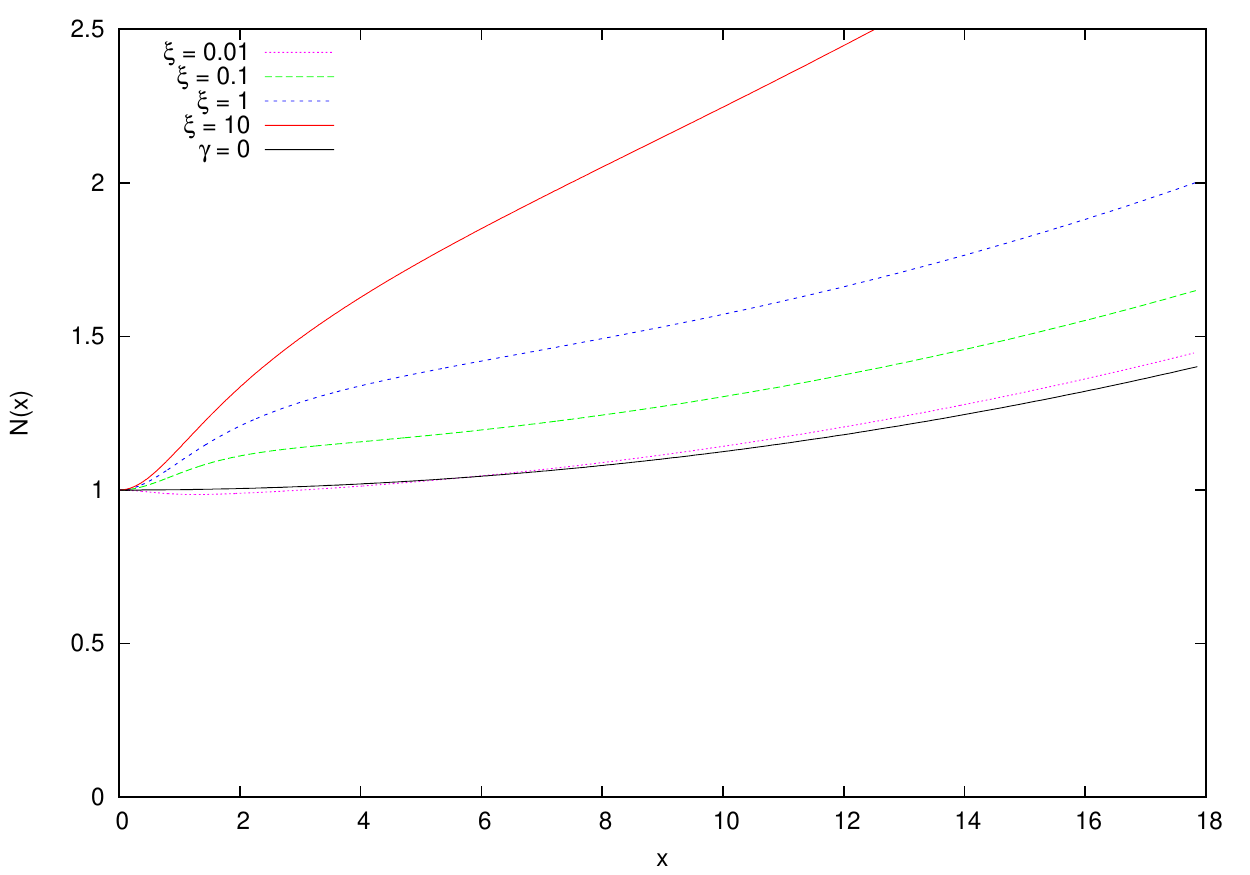}
\end{tabular}
\caption{The metric function $N(x)$ in the presence of a cosmological constant, $\bar{\Lambda}$. At left, $\bar{\Lambda} > 0$. At right, $\bar{\Lambda} < 0$. In both cases, the profiles are plotted for several values of the parameter $\xi$, plus a plot of the same system in vacuum ($\gamma = 0$). The parameters used for the cosmic string were $\alpha = 0.8, \beta_2 = 2.0, \beta_3 = 1.0$.}
\label{fig:AntiDeSitter}
\end{figure}

In figure (\ref{fig:AntiDeSitter}) we plot a graph for several profiles of the function $N(x)$ of the metric, for fixed $\gamma = 0.5, \alpha = 0.8, \beta_2 = 2.0$ and $\beta_3 = 1.0$. The variation of the metric function along such parameters has been already studied in \cite{Santo:2015xma} and we will not repeat the results here. We are interested to compare how the metric function changes as we move from Einstein's gravity to the Starobinsky model of gravity. At left, we show the profiles for de Sitter, and at right we show the profiles for anti-de Sitter. In both cases, we are using the parameter $\bar{\Lambda} = 0.005$, and in both cases we show in black the profiles for the case where $\gamma = 0$, i.e, when the gravitational field decouples from the matter.

For de Sitter (left), we can note that the presence of matter decreases the value of the cosmological horizon $x_0$ for small values of the parameter $\xi$ (or, in Einstein's gravity). In the presence of the $R^2$ correction, as we increase the parameter $\xi$, the value for the cosmological horizon approaches the value it would have in absence of matter. For larger values of the parameter $\xi$, $x_0$ with matter becomes bigger than $x_0$ in the absence of matter. This is certainly not a trivial result, and it is a feature of the extended theory. 

For anti-de Sitter (right), the increase of the parameter $\xi$ affects the metric function $N(x)$ as showed, but is hard to qualify  
the effects of such modification. What we can say is that, in the case of a negative cosmological constant, the effect of the $R^2$ term appears to be different from the effect of the same term in an asymptotically flat spacetime. To clarify this sentence: In anti-de Sitter, the effect of the correction term is not to attenuate the matter contribution to geometrical effects.    

\section{Conclusions}

In this paper we have studied a model for a non-Abelian cosmic string in the Starobinsky model of gravity. Such model can be recast as an $f(R)$ theory of gravity, where $f(R) = R^2 + \zeta R^2$. The main idea of the paper is to study how the topological and geometrical effects due to a cosmic string changes as we change the parameter $\zeta$ or, more precisely, its dimensionless counterpart $\xi$. 

The most remarkable feature of cosmic strings in an asymptotically flat spacetimes is the generation of an angular deficit. Such angular deficit can, in principle, be observed by its influence on light rays coming from astrophysical sources. Despite observational efforts, to the present moment not a single candidate for a cosmic string has been identified. One possible explanation for such a failure can be that our gravitational theory is not Einstein's gravity, but a theory as the Starobinsky model of gravity. We have shown that, for a non-Abelian model of a cosmic string, the angular deficit generated by such a source in this $f(R)$ gravity is smaller than what would be the value in pure Einstein's gravity. We have show this feature for several string profiles, generated by different string parameters.

We also studied the behaviour of the metric function $N(x)$ in the presence of a cosmological constant, both positive and negative. For a positive cosmological constant (de Sitter), we have found that the  cosmological horizon changes from smaller to larger as we increase the parameter $\xi$. For the case of an anti-de Sitter spacetime, the profile of the metric function $N(x)$ has also been plotted, but its harder to present definitive conclusions, since anti-de Sitter has not a fiducial parameter that we can quantify and analyze.  

\section*{Acknowledgments}
This work was supported by Brazilian agency CNPq (Conselho Nacional de Desenvolvimento Cient\'ifico e Tecnol\'ogico). ERBM is partially supported by research project $N_o$ $313137/2014-5$. VBB is partially supported by research project $N_o$ $305835/2016-5$.

\end{document}